\documentclass{elsart}
\usepackage{graphicx}

\def\ba{\begin{eqnarray}}
\def\ea{\end{eqnarray}}

\begin{document}

\begin{frontmatter}

\title{\bf The realistic QCD equation of state in relativistic heavy-ion collisions and the early Universe \thanksref{grant}} 
\thanks[grant]{e-mail: Wojciech.Florkowski@ifj.edu.pl \\.}

\author[ujk,ifj]{Wojciech Florkowski} 

\address[ujk]{Institute of Physics, Jan Kochanowski University,
ul.~\'Swi\c{e}tokrzyska 15, 25-406~Kielce, Poland} 
\address[ifj]{The H. Niewodnicza\'nski Institute of Nuclear Physics, Polish Academy of Sciences, ul. Radzikowskiego 152, 31-342 Krak\'ow, Poland}

\begin{abstract} 
The realistic equation of state of strongly interacting matter, that has been successfully applied in the recent hydrodynamic studies of hadron production in relativistic heavy-ion collisions at RHIC, is used in the Friedmann equation to determine the precise time evolution of thermodynamic parameters in the early Universe. A comparison with the results obtained with simple ideal-gas equations of state is made. The realistic equation of state describes a crossover rather than the first-order phase transition between the quark-gluon plasma and hadronic matter. Our numerical calculations show that small inhomogeneities of strongly interacting matter in the early Universe are moderately damped during such crossover. 
\end{abstract}
\end{frontmatter}
\vspace{-7mm} PACS: 25.75.Nq, 98.80.Cq

\section{Introduction}
\label{sect:intro}

One expects that the studies of relativistic heavy-ion collisions may bring us information about the properties of matter in the early Universe (at the time when its age was about 10 microseconds and its temperature was about 200 MeV). In this paper we give an explicit example showing how such expectations may be realized in practice. Our main point is that the theoretical analysis of the hadronic observables collected at RHIC (Relativistic Heavy Ion Collider at BNL) pinned down the form of the equation of state (EOS) of strongly interacting matter. The use of this EOS in the Friedmann equation allows us to determine the precise time evolution of the thermodynamic parameters in the early Universe. 

Until recently, the soft hadronic observables measured at RHIC have been described with the help of the hydrodynamic models which incorporated the first-order phase transition between the quark-gluon plasma and hadronic matter \cite{Kolb:2003dz,Huovinen:2003fa}. The main success of this approach was the correct reproduction of the hadron transverse-momentum spectra and the elliptic-flow coefficient $v_2$. On the other hand, the hydrodynamic models using the first-order phase transition had severe problems with the reproduction of pionic correlation functions. The latter contain the information about the space-time dimensions of hadronic systems formed in the collisions. Thus, the evident {\it successes} in the reproduction of the hadron {\it momentum distributions} were obscured by the {\it failures} connected with the analysis of the {\it space-time dimensions} of the produced system. 

The large part of those difficulties, known in the literature as the HBT puzzle, was removed by the use of the EOS with the crossover phase transition \cite{Broniowski:2008vp,Pratt:2008qv}. The crossover phase transition is not a genuine phase transition --- all thermodynamic parameters change smoothly as functions of the temperature, however, such changes may be quite sudden in a narrow temperature range. The crossover phase transition {\it without} a distinct {\it soft point} (i.e., without a thermodynamic region where the ratio of pressure and energy density is very small and the sound velocity possibly drops to zero) leads to a faster time development of the system and helps to reproduce the correlation data. 

The use of the simple bag EOS (leading to the first-order phase transition) in the hydrodynamic equations describing heavy-ion collisions is very much reminiscent of the situation in the cosmological calculations where, in most cases, the strongly-interacting matter is treated in the same naive way. Therefore, it is interesting to search for the effects of using the realistic QCD EOS in the Friedmann equation. In particular, the results of the lattice simulations of QCD with physical quark masses, which suggest the crossover rather than the first-order phase transition, should be incorporated in such analyses. The work in this direction has been recently done in Ref. \cite{Sassi:2009tr} where the lattice results from \cite{Cheng:2007jq} were used. 

Our present approach also uses the lattice results but differs from Ref. \cite{Sassi:2009tr} in several important aspects. Firstly, we use different lattice results as an input \cite{Aoki:2005vt}. Secondly, our EOS is not based exclusively on the lattice results. At lower temperatures we use the realistic hadron-gas model, which turned out to be very successful in the thermodynamic analyses of hadron production in heavy-ion collisions \cite{Florkowski:2001fp}. The results of the hadron-gas model at lower temperatures are combined with the lattice results at higher temperatures in the thermodynamically  consistent way described in \cite{Chojnacki:2007jc}. However, the most attractive feature of our EOS is the fact that it was successfully implemented in 2+1 \cite{Chojnacki:2007rq} and 3+1 \cite{Bozek:2009ty} hydrodynamic codes describing heavy-ion collisions. Thus, in our case the knowledge from heavy-ion physics is transferred directly to cosmology. We note that an EOS sharing very similar features to ours has been very recently constructed in Ref. \cite{Huovinen:2009yb}. 

In the second part of the paper we consider the time development of the energy density fluctuations using the theory developed by Schmid, Schwarz, and Widerin \cite{Schmid:1996qd,Schmid:1998mx}. By doing explicit numerical calculations we show that small inhomogeneities of the strongly interacting matter are moderately (by about 30\%) damped during the realistic crossover transition. 

Altogether, our results indicate that there are small chances for observation exotic phenomena connected usually with the first order phase transitions (the latter are discussed, for example, in \cite{Boyanovsky:2006bf}). 

\section{Equations of state for strongly interacting matter}
\label{sect:eoss}

\begin{figure}[t]
\begin{center}
\includegraphics[angle=0,width=0.65\textwidth]{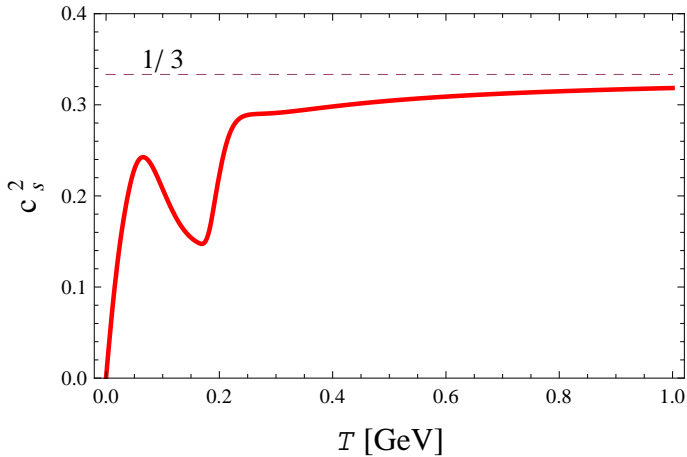}
\end{center}
\caption{\small Solid line: temperature dependence of the sound velocity for the realistic QCD EOS constructed in \cite{Chojnacki:2007jc} and used in this paper. Dashed line: the limit for massless and non-interacting particles.}
\label{fig:cs2}
\end{figure}

The QCD EOS used in this paper was introduced and discussed in greater detail in Ref. \cite{Chojnacki:2007jc}. As mentioned above, it interpolates between the hadron-gas results and the lattice results. All thermodynamic properties follow from the temperature dependence of the sound velocity $c_s^2$ depicted in Fig.~\ref{fig:cs2}. The smooth dependence of the sound velocity on the temperature indicates that there is no genuine phase transition in the system. On the other hand,  $c_s^2(T)$ has a characteristic minimum at $T \sim$ 170 MeV. This behavior reflects the presence of the sharp crossover in this region, i.e., the region where the energy density increases strongly within a narrow range of temperature.  We stress that the values of $c_s^2$ differ from 1/3. This indicates that the matter under consideration does not behave like an ideal gas of massless particles.

\begin{figure}[t]
\begin{center}
\includegraphics[angle=0,width=0.65\textwidth]{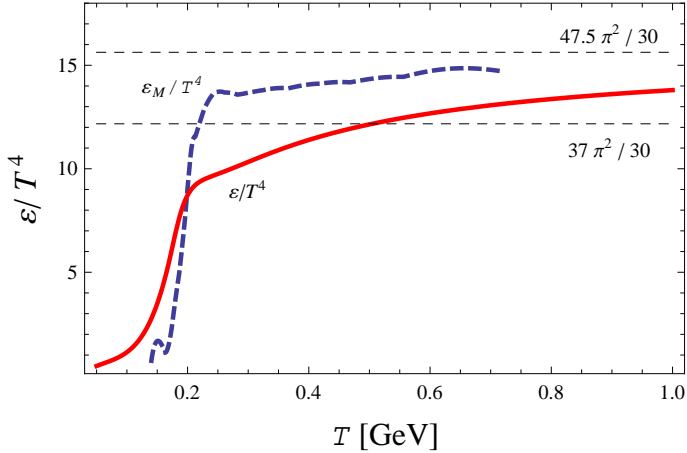}
\end{center}
\caption{\small Solid thick line: the energy density $\varepsilon$  obtained from the function ${c}_s^2(T)$ shown in Fig.~\ref{fig:cs2} and scaled by $T^4$. Dashed thick line: the energy density $\varepsilon_M$ used in Ref. \cite{Sassi:2009tr}, also scaled by $T^4$. Thin dashed lines: the Stefan-Boltzmann limits for the ratio $\varepsilon/T^4$ for the bag EOS with 2 (lower line) and 3 (upper line) flavors. }
\label{fig:sM}
\end{figure}

Various thermodynamic quantities characterizing our QCD EOS may be obtained from the thermodynamic identities valid at zero baryon chemical potential
\begin{eqnarray}
\varepsilon + P = T \sigma, \quad d\varepsilon = Td\sigma, \quad dP = \sigma dT, \quad
c_s^2 = \frac{dP}{d\varepsilon} = \frac{\sigma dT}{T d\sigma}.
\end{eqnarray}
Here $\varepsilon$, $P$, $T$, and $\sigma$ are the energy density, pressure, temperature, and entropy density, respectively. In Fig.~\ref{fig:sM} we show the energy density scaled by $T^4$ (solid line), that was obtained directly from the function $c_s^2(T)$ depicted in Fig.~\ref{fig:cs2}. 

For comparison, we also show the energy density $\varepsilon_M$ used by the Milan group in Ref. \cite{Sassi:2009tr}. The function $\varepsilon_M(T)$ is again scaled by $T^4$ (dashed line). We observe that our values of $\varepsilon/T^4$ are larger at lower temperatures and smaller at higher temperatures. Larger values of $\varepsilon/T^4$ at lower temperatures are connected with the inclusion of all known resonances in the hadron-gas model. On the other hand, smaller values of $\varepsilon/T^4$ at higher temperatures indicate stronger interactions in the plasma phase. 

In order to compare the results obtained with the realistic QCD EOS with the results obtained with the bag EOS we introduce the following definitions
\begin{eqnarray}
{\varepsilon}_{\rm qgp} &=&  g_{\rm qgp} \frac{\pi^2}{30} T^4+B, 
\quad P_{\rm qgp} = g_{\rm qgp} \frac{\pi^2}{90} T^4-B, \quad \hbox{if} 
\quad  T > T_c, \label{Pqgp} \\
{\varepsilon}_{\pi} &=&
 g_\pi \frac{\pi^2}{30} T^4, \hspace{1.75cm} P_{\pi} = g_\pi \frac{\pi^2}{90} T^4, \quad \hspace{1.25cm}  \hbox{if}\quad T < T_c. \label{Ppi}
 \end{eqnarray}
Here $B$ is the bag constant and $T_c$ is the critical temperature defined by the formula
\begin{equation}
T_c = \left[\frac{90 B}{(g_{\rm qgp}-g_\pi)\pi^2} \right]^{1/4}.
\label{Tc}
\end{equation}

Equation (\ref{Pqgp}) describes the energy density and pressure of the weakly interacting quark-gluon plasma, whereas Eq. (\ref{Ppi}) describes the energy density and pressure of the massless pion gas. For two quark flavors  \mbox{$g_{\rm qgp} =37$}, and for three flavors \mbox{$g_{\rm qgp}=47.5$}. 

In the numerical calculations presented below we use \mbox{$g_{\rm qgp} =37$} and consider two values of the bag constant:  $B = (235\, \hbox{MeV})^4$ and $B = (165\, \hbox{MeV})^4$. They correspond to two values of the critical temperature: \mbox{$T_c$ = 169 MeV}  and \mbox{$T_c$ = 119 MeV}.

We stress that the use of the bag EOS as a reference point for the calculations with the realistic QCD EOS is ambiguous. For example, we easily find that the bag EOS yields the Stefan-Boltzmann limit $\varepsilon_{\rm qgp}/T^4 = 37 \pi^2/30$ for two flavors and $\varepsilon_{\rm qgp}/T^4 = 47.5 \pi^2/30$ for three flavors. Figure \ref{fig:sM} shows that the realistic value of $\varepsilon/T^4$ does not reach any of those limits. The difference indicates that the interactions in the plasma are non-negligible at the temperatures as high as \mbox{1 GeV}. 

Below, we consider the time evolution of the thermodynamic quantities governed by the Friedmann equation. The initial time corresponds to the moment when \mbox{$T = T_0$ = 500 MeV} and \mbox{$\varepsilon(T_0)$ = 100 GeV/fm$^3$}. This value is close to the energy density obtained at the same temperature with the bag EOS with two flavors. This {\it coincidence} suggests that it is useful to consider the bag EOS with two flavors as a reference EOS for $ T \leq 500$ MeV.

\section{Inclusion of the electro-weak sector}

In the temperature range considered by us in the Friedmann equation, one should include also the energy density and pressure of the particles from the electro-weak sector. Typically they are treated as massless particles, whose effective number of degrees of freedom is $g_{\rm ew} = 14.45$ \cite{Yagi:2005yb}. Thus we write
\begin{eqnarray}
\varepsilon_R = \varepsilon + \varepsilon_{\rm ew}, \quad P_R = P + P_{\rm ew},
\label{epsandP}
\end{eqnarray}
where $\varepsilon_R$ and $P_R$ are the total energy density and pressure used in the Friedmann equation and 
\begin{eqnarray}
\varepsilon_{\rm ew} = g_{\rm ew} \frac{\pi^2}{30} T^4, \quad P_{\rm ew} = g_{\rm ew} \frac{\pi^2}{90} T^4, \quad \sigma_{\rm ew} = \frac{4 \varepsilon_{\rm ew}}{3 T}.
\label{epsandTew}
\end{eqnarray}

\begin{figure}[t]
\begin{center}
\includegraphics[angle=0,width=0.65\textwidth]{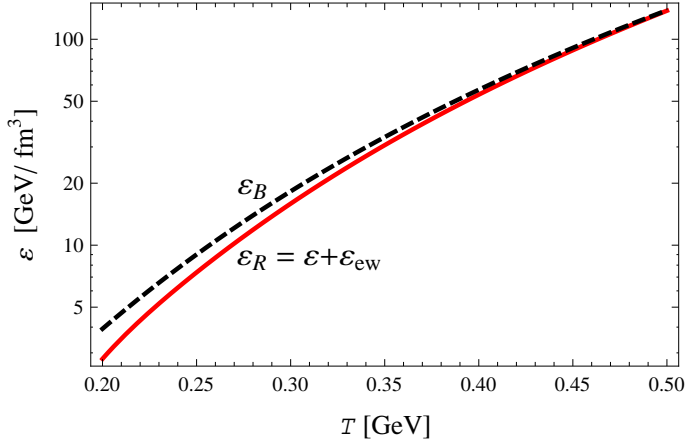}
\end{center}
\caption{\small Temperature dependence of the total energy density $\varepsilon_R = {\varepsilon} + \varepsilon_{\rm ew}$ and the reference energy density $\varepsilon_B$ obtained with the bag EOS. }
\label{fig:epsRB}
\end{figure}

In the similar way we construct the modified bag EOS
\begin{eqnarray}
{\varepsilon}_{B} &=&  g_{1} \frac{\pi^2}{30} T^4+B, 
\quad P_{B} = g_{1} \frac{\pi^2}{90} T^4-B, 
\hspace{1.1cm} \hbox{if} \quad  T > T_c, \nonumber \\
{\varepsilon}_{B} &=& g_2 \frac{\pi^2}{30} T^4, \hspace{1.3cm} P_{B} = g_2 \frac{\pi^2}{90} T^4, \quad \hspace{1.55cm}  \hbox{if}\quad T < T_c, \label{PB}
 \end{eqnarray}
where \mbox{$g_1=37+g_{\rm ew}=51.25$} and \mbox{$g_2=3+g_{\rm ew}=17.25$}. 

Since $g_{\rm qgp}-g_\pi = g_1-g_2$, the value of the critical temperature for fixed $B$ remains the same as in the case without the electro-weak sector. We note that we use now the same subscript to denote the thermodynamic variables above and below $T_c$. The temperature dependence of the functions $\varepsilon_R(T)$ and $\varepsilon_B(T)$ in the temperature range \mbox{0.2 GeV $\leq T \leq$ 0.5 GeV} is shown in Fig. \ref{fig:epsRB}. We observe that the two functions converge in the region \mbox{0.45 GeV $\leq T \leq$ 0.50 GeV}.

\section{Solutions of the Friedman equation}

Assuming that the expansion of the Universe is isentropic, the Friedmann equation in our case takes the form
\begin{equation}
\frac{d\varepsilon_R}{dt} = -3 \sqrt{\frac{8\pi G \varepsilon_R}{3}} 
(\varepsilon_R + P_R).
\label{frid1}
\end{equation}
In the next step we use Eq. (\ref{epsandP}) and switch to the temperature as the independent variable. In this way we get
\begin{equation}
\left[c_s^{-2} \sigma + 3 \sigma_{\rm ew} \right] \frac{dT_R}{dt} 
= -3 \sqrt{\frac{8\pi G (\varepsilon+\varepsilon_{\rm ew})}{3}} 
(\varepsilon+\varepsilon_{\rm ew} + P + P_{\rm ew}).
\label{frid2}
\end{equation}
We note that $c_s^{-2}$, $\sigma$, $\varepsilon$, and $P$ in Eq. (\ref{frid2}) characterize the realistic QCD EOS, while $\sigma_{\rm ew}$, $\varepsilon_{\rm ew}$, and $P_{\rm ew}$ are defined by Eq. (\ref{epsandTew}). The temperature obtained from Eq. (\ref{frid2}) has a subscript $R$ in order to distinguish it from the temperature profile $T_B$ obtained for the bag EOS \footnote{The reader may connect the subscript $R$ with the {\it realistic} or RHIC equation of state.  }. 

\begin{figure}[t]
\begin{center}
\includegraphics[angle=0,width=0.65\textwidth]{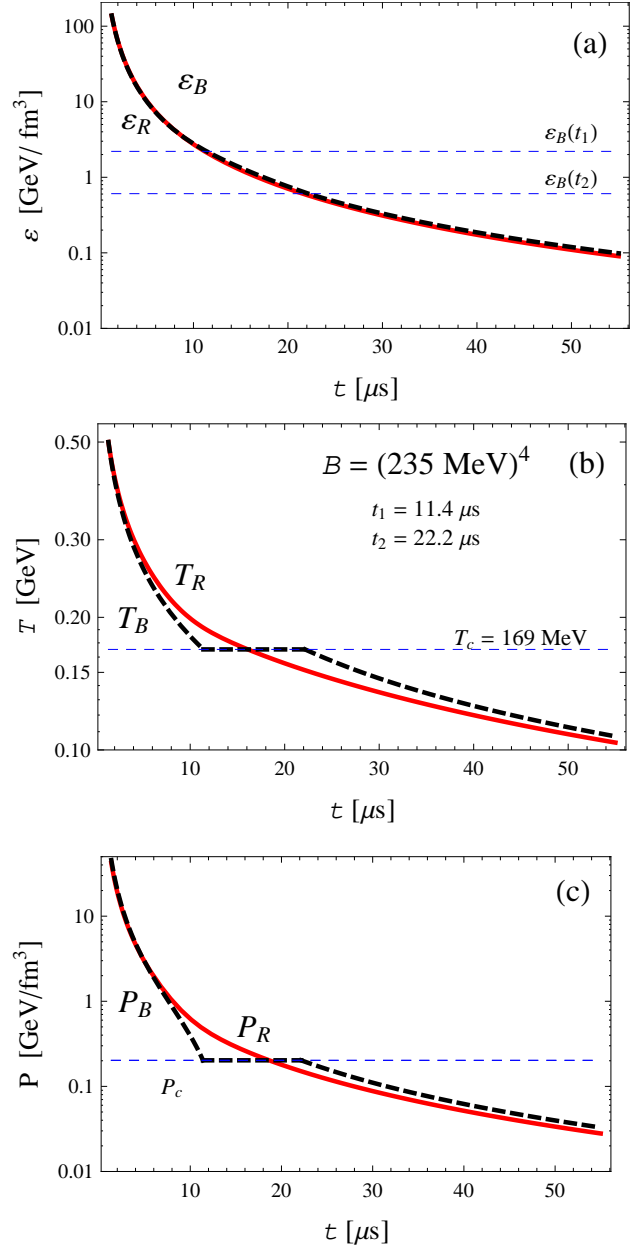}
\end{center}
\caption{\small The energy density $\varepsilon_R$ {\bf (a)}, temperature $T_R$ {\bf (b)}, and pressure $P_R$ {\bf (c)}, obtained by solving the Friedmann equation with the realistic QCD EOS with the initial condition $T_R$ = 500 MeV at $t_0$ = 1.35 $\mu s$ (solid lines). The dashed lines show the results obtained for the bag EOS with \mbox{$B$= (235 MeV)$^4$} and the same initial condition, $T_B(t_0)=T_R(t_0)$. The initial and final times for the reference first-order phase transition are denoted by $t_1$ and $t_2$, respectively. }
\label{fig:B235}
\end{figure}

The initial condition for solving Eq. (\ref{frid2}) has the form
\begin{equation}
T_R(t_0) = 500 \, \hbox{MeV}.
\label{T0}
\end{equation}
The corresponding initial energy density 
\begin{equation}
\varepsilon_{R,0} = \varepsilon(T_R(t_0)) + \varepsilon_{\rm ew}(T_R(t_0)) = 138 \, \hbox{GeV/fm$^3$}.
\label{eps0P0}
\end{equation}
Assuming that the evolution for $t < t_0$ is dominated by matter with the radiation-like EOS we set 
\begin{equation}
t_0 = \sqrt{\frac{3}{32 \pi G \varepsilon_{R,0}}} = 1.35\, \mu s.
\label{t0}
\end{equation}

\begin{figure}[t]
\begin{center}
\includegraphics[angle=0,width=0.65\textwidth]{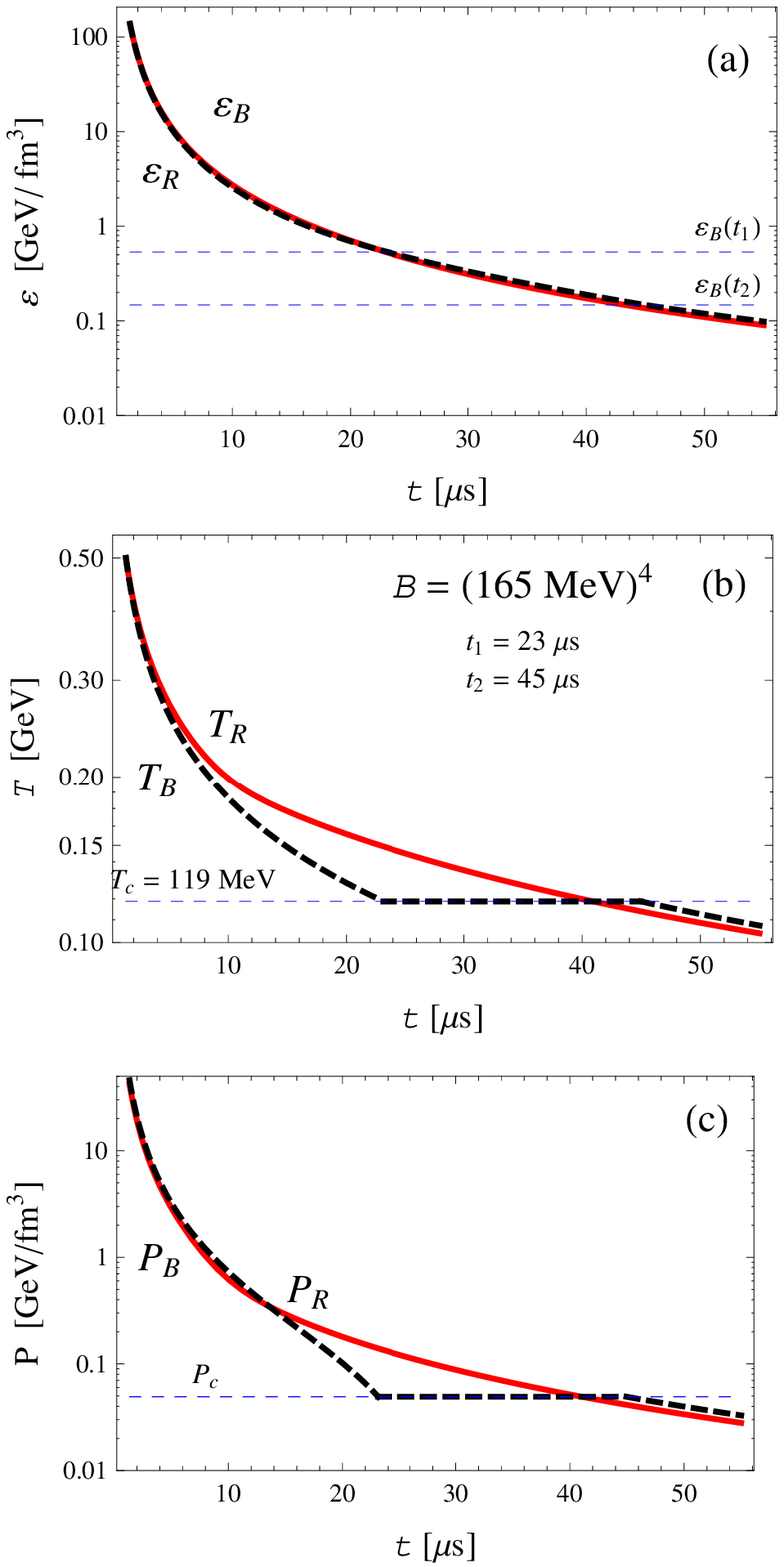}
\end{center}
\caption{\small The same as Fig. \ref{fig:B235} but with the reference bag calculation done for \mbox{$B$= (165 MeV)$^4$}. }
\label{fig:B165}
\end{figure}

The numerical solution of Eq. (\ref{frid2}), i.e., the function $T_R(t)$ as well as the functions $\varepsilon_R(t)$ and $P_R(t)$ are shown in Figs. \ref{fig:B235} and \ref{fig:B165} (thick solid lines). The results obtained with the realistic QCD EOS are compared with the results obtained with the bag EOS (thick dashed lines) and the same initial conditions. The two figures differ by the value of the bag constant used in the reference calculation: $B = (235\, \hbox{MeV})^4$ and $B = (165\, \hbox{MeV})^4$ in Figs. \ref{fig:B235} and \ref{fig:B165}, respectively. 

The time changes of the functions $\varepsilon_R(t)$ and $\varepsilon_B(t)$ shown in Fig. \ref{fig:B235} are very much similar. On the other hand, the changes of the temperature and pressure are quite different in the region where the system governed by the bag EOS passes through the phase transition. For large values of time, the functions $T_R$ and $P_R$ become again close to $T_B$ and $P_B$.

The value of the bag constant used in the calculation shown in Fig. \ref{fig:B235} implies \mbox{$T_c$ = 169 MeV}. This value agrees with the critical temperature used in the realistic QCD EOS. Strictly speaking, the realistic crossover phase transition does not allow for the unique definition of the critical temperature. Nevertheless, in our approach we  associate $T_c$ of the crossover with the local minimum of the sound velocity function shown in Fig. \ref{fig:cs2}, which is situated close to \mbox{$T$ = 170 MeV}.

In Fig. \ref{fig:B165} we compare the same results obtained with the realistic QCD EOS with the reference calculations based on the bag model with $B = (165\, \hbox{MeV})^4$. In this case the value of the bag constant corresponds to \mbox{$T_c$ = 119 MeV}. Clearly, the agreement between the two results is much poorer. 

\section{Scale factor and conformal time}

\begin{figure}[t]
\begin{center}
\includegraphics[angle=0,width=0.65\textwidth]{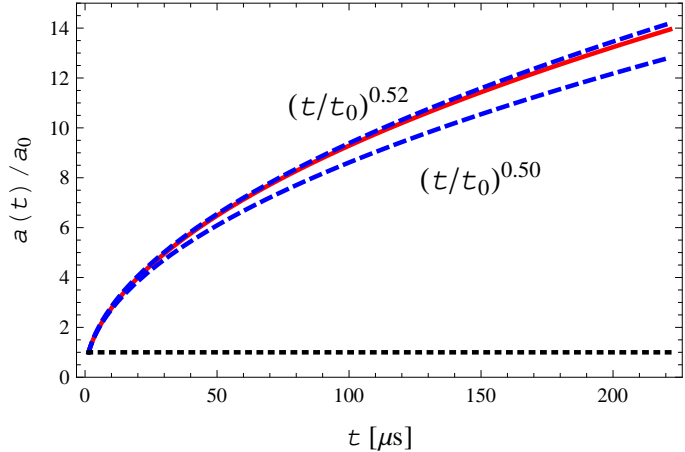}
\end{center}
\caption{\small Time evolution of the scale factor $a(t)$ (solid line) compared with two power-law expressions: $(t/t_0)^{0.50}$ (lower dashed line) and  $(t/t_0)^{0.52}$ (upper dashed line). }
\label{fig:a}
\end{figure}

The time evolution of the scale factor $a(t)$ is given by the formula
\begin{equation}
a(t) = a(t_0) \exp\left[ \, \,
\int_{t_0}^t \sqrt{\frac{8 \pi G \varepsilon_R(t')}{3}} dt' 
\right].
\label{fig:at}
\end{equation}
In Fig. \ref{fig:a} we show the time dependence of $a(t)/a(t_0)$ obtained with our energy density function $\varepsilon_R$ (solid line). For comparison, we also show the two power-law functions   $(t/t_0)^p$, where $p=0.50$ (lower dashed line) and  $p=0.52$ (upper dashed line). Clearly, the use of the realistic EOS changes slightly the standard massless-gas behavior corresponding to $p=0.5$. 

In the following we use the approximation of $a(t)/a(t_0)$ by the function of the form $(t/t_0)^p$ and in the numerical calculations we set $p=0.517$  \footnote{The value $p=0.517$ follows from the power-law fit done in the whole time interval between  \mbox{$t$ = 1.35 $\mu$s}  and \mbox{$t$ = 220 $\mu$s}. For \mbox{$t < $  10 $\mu$s} the function $a(t)/a(t_0)$ is much better described by the formula $(t/t_0)^{0.5}$, however, the presence of the phase transition modifies the effective power-law behavior at larger times. We note that the change of $p$ from 0.5 to 0.517 results in the 12\% change of $a$ at \mbox{$t$ = 220 $\mu$s}.}. In this case, the Hubble constant is
\begin{equation}
H = \frac{da}{a \, dt} = \frac{p}{t}.
\label{H}
\end{equation}
For further applications it is useful to introduce the conformal time $\eta$ defined by the relation
\begin{equation}
\frac{d\eta}{dt} = \frac{1}{a},
\label{eta}
\end{equation}
and the quantity ${\cal H}$ that is related to the Hubble constant, 
\begin{equation}
{\cal H} = a H = \frac{da}{a \, d\eta}.
\label{Hcal}
\end{equation}
Simple calculations lead to 
\begin{equation}
\eta = \frac{t_0^p \, t^{1-p}}{(1-p) \, a_0}, 
\quad {\cal H} = \frac{p}{(1-p)\,\eta}.
\label{etaandHcal}
\end{equation}
We further define the parameter $k_0$ and the variable $x$,
\begin{equation}
a_0 = \frac{k_0 t_0}{1-p}, \quad x = k_0 \eta = \left( \frac{t}{t_0} \right)^{1-p}.
\label{k0x}
\end{equation}

\section{Cosmological inhomogeneities at the QCD phase transition}

The character of the QCD phase transition affects not only the time dependence of average thermodynamic  quantities but it has also influence on the primordial density fluctuations in the early Universe. If the QCD phase transition was of the first order, with the  sound velocity vanishing at the critical temperature, the density fluctuations would develop large peaks and dips \cite{Schmid:1996qd,Schmid:1998mx}. For the crossover transition we do not expect any spectacular phenomena, however, it is interesting to analyze the development of fluctuations in the realistic background analyzed in the previous Sections and to check if the drop of the sound velocity at $T \sim$ 170 MeV may have any noticeable effects on the  inhomogeneities. 

Below we study the system of equations introduced by Schmid, Schwarz, and Widerin \cite{Schmid:1996qd,Schmid:1998mx}. They read 
\begin{eqnarray}
\frac{1}{\cal H} \delta^\prime + 3 (c_s^2 - w) \delta &=& \frac{k}{\cal H} \psi - 3 (1+w) \, \alpha, 
\label{eq1} \\
\frac{1}{\cal H} \delta^\prime_{\rm ew}  &=& \frac{k}{\cal H} \psi_{\rm ew} - 4 \alpha, 
\label{eq1ew} \\
\frac{1}{\cal H} \psi^\prime + (1 - 3 w) \psi &=& -c_s^2 \frac{k}{\cal H} \delta - (1+w) \frac{k}{\cal H}  \alpha,
\label{eq2} \\
\frac{1}{\cal H} \psi^\prime_{\rm ew} &=& - \frac{k}{3 \cal H} \delta_{\rm ew} -  \frac{4 k}{3 \cal H}  \, \alpha,
\label{eq2ew} \\
\left[ \left( \frac{k}{\cal H} \right)^2 + \frac{9}{2} \left(1+w_R\right) \right] \alpha
&=& -\frac{3}{2} \left( 1+ 3 c_{sR}^2 \right) \, \delta_R. \label{eq3}
\end{eqnarray}
Equations (\ref{eq1}) and (\ref{eq1ew}) follow from the energy-momentum conservation, Eqs. (\ref{eq2}) and (\ref{eq2ew}) from the 3-divergence of the Euler equation of general relativity, and Eq. (\ref{eq3}) from the Einstein $R^0_0$-equation. The prime  denotes the derivative with respect to the conformal time $\eta$. The quantities $\delta$ describe the energy density fluctuations (density contrasts), $\psi's$ are related to the fluid velocities (peculiar velocities), and $\alpha$ defines the correction to the temporal part of the metric tensor (lapse function). The function $w$ is the ratio of pressure and energy density for the background evolution, and $c_s$ is the sound velocity of the background (in our case we use here the results of solving the Friedmann equation presented in the previous Sections). 

Equations (\ref{eq1}) and (\ref{eq2}) refer to the QCD matter, whereas Eqs. (\ref{eq1ew}) and (\ref{eq2ew}) refer to the electro-weak sector. On the other hand, Eq. (\ref{eq3}) refers to the two components, hence, in this case we have
\begin{eqnarray}
w_R &=& \frac{P_R}{\varepsilon_R} 
= \frac{P + P_{\rm ew}}{\varepsilon + \varepsilon_{\rm ew}}, \nonumber \\
c_{sR}^2 &=& \frac{dP_R}{d\varepsilon_R} 
= \frac{dP + dP_{\rm ew}}{d\varepsilon + d\varepsilon_{\rm ew}}
= \frac{\sigma + \sigma_{\rm ew}}{c_s^{-2} \, \sigma + 3 \sigma_{\rm ew}},
\end{eqnarray}
and
\begin{eqnarray}
\delta_R &=& \frac{\delta \varepsilon_R}{\varepsilon_R}
= \frac{\delta \varepsilon  + \delta \varepsilon_{\rm ew}}{\varepsilon + \varepsilon_{\rm ew}} = \frac{\delta   + \delta_{\rm ew} (\varepsilon_{\rm ew}/\varepsilon) }{1 + (\varepsilon_{\rm ew}/\varepsilon)}.
\end{eqnarray}
The functions $P, P_{\rm ew}, \varepsilon, \varepsilon_{\rm ew}, \sigma$ and $\sigma_{\rm ew}$ are known if the temperature dependence on time is established by Eq. (\ref{frid2}).

The parameter $k$ in Eqs. (\ref{eq1})--(\ref{eq3}) is the comoving wavelength, $k = k_{\rm phys} \,a$. In the numerical calculations we switch from the conformal time $\eta$ to the variable $x = k_0 \eta$. In this case, using the variable $\xi = k/k_0$, see Eq. (\ref{k0x}), we may write
\begin{eqnarray}
\frac{1}{\cal H} \delta^\prime &=& \frac{1-p}{p} \eta \frac{d\delta}{d\eta} 
= \frac{1-p}{p} x \frac{d\delta}{dx} \equiv q \, x \, {\dot \delta}, \nonumber \\
\frac{k}{\cal H} &=& \frac{1-p}{p} \eta k \equiv q  \xi x.
\label{prime-dot}
\end{eqnarray}
The dot in (\ref{prime-dot}) denotes differentiation with respect to $x$. Using substitutions of the type (\ref{prime-dot}) in Eqs. (\ref{eq1})--(\ref{eq3})  we obtain a system of 5 coupled ordinary differential equations for 5 unknown functions: $\delta$, $\delta_{\rm ew}$, $\psi$, $\psi_{\rm ew}$, and $\alpha$:
\begin{eqnarray}
q x \, {\dot\delta} + 3 (c_s^2 - w) \,  \delta &=& q \xi x \,  \psi - 3 (1+w) \, \alpha, 
\label{eq1x} \nonumber \\
q x \, {\dot \delta}_{\rm ew}  &=& q \xi x \,  \psi_{\rm ew} - 4 \, \alpha, 
\label{eq1ewx} \nonumber \\
q x \, {\dot \psi} + (1 - 3 w) \, \psi &=& - q \xi x \left[ c_s^2  \delta + (1+w)  \alpha \right],
\label{eq2x} \nonumber  \\
 {\dot \psi}_{\rm ew} &=& - \frac{\xi}{3} \left(   \delta_{\rm ew} + 4  \, \alpha \right),
\label{eq2ewx} \nonumber \\
\left[ \left( q \xi x  \right)^2 + \frac{9}{2} \left(1+w_R\right) \right] \alpha
&=& -\frac{3}{2} \left( 1+ 3 c_{sR}^2 \right) \, \delta_R. \label{eq3x}
\end{eqnarray}

We analyze the evolution of inhomogeneities in the time interval between  \mbox{$t$ = 1.35 $\mu$s}  and \mbox{$t$ = 220 $\mu$s}. Since $x = (t/t_0)^{(1-p)}$, the starting value of $x$ is 1 and the final value of $x$ is about 12. During this evolution, the temperature of the system drops to 53 MeV, thus we explore the whole temperature range where the sound velocity has a minimum, see Fig. \ref{fig:cs2}.

\begin{figure}[t]
\begin{center}
\includegraphics[angle=0,width=0.65\textwidth]{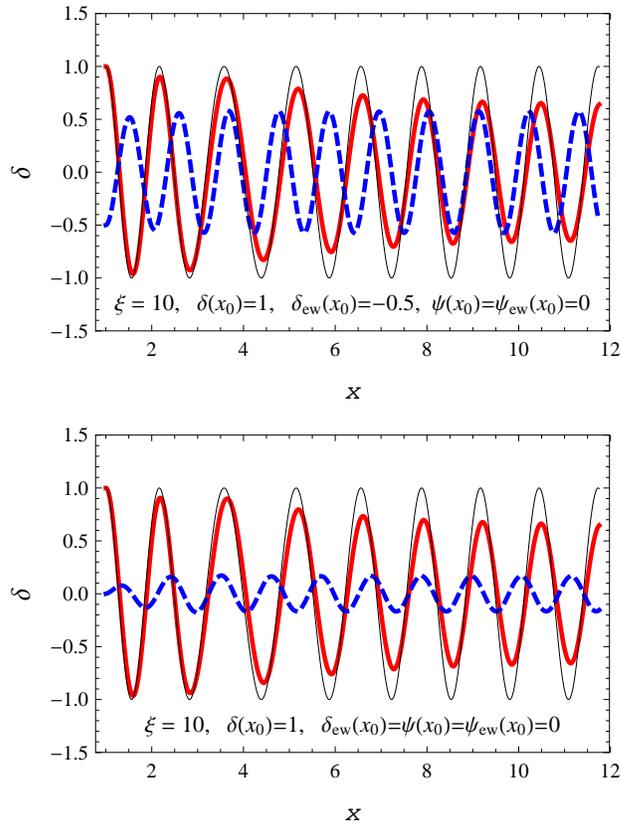}
\end{center}
\caption{\small Time evolution of the density contrasts $\delta$ (solid line) and $\delta_{\rm ew}$ (dashed line) for $\xi=10$. The values of the initial conditions are given in the Figures.}
\label{fig:xi10ab}
\end{figure}

\begin{figure}[t]
\begin{center}
\includegraphics[angle=0,width=0.65\textwidth]{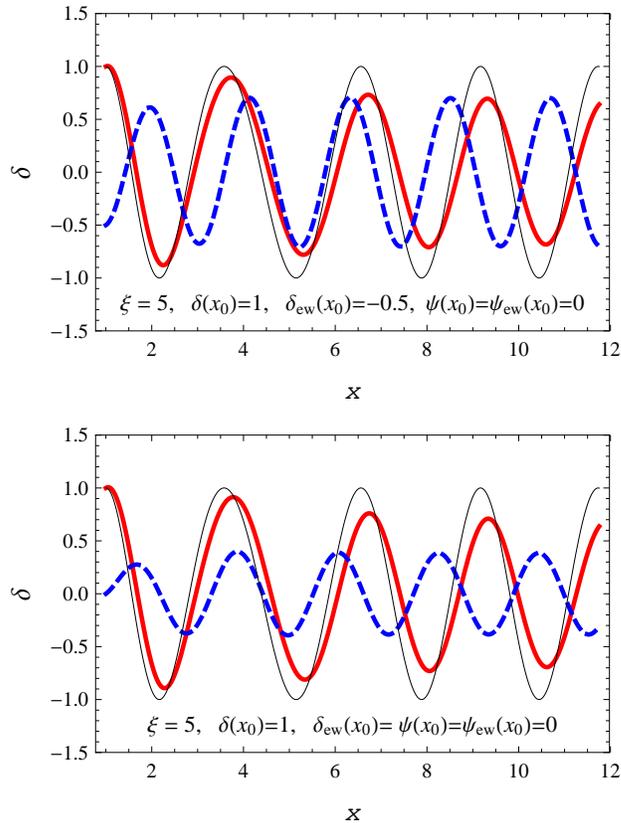}
\end{center}
\caption{\small The same as Fig. \ref{fig:xi10ab} but for $\xi=5$.  }
\label{fig:xi5ab}
\end{figure}

In the numerical calculations we consider the subhorizon modes that satisfy the condition \mbox{$k/{\cal H} \gg 1$}~\footnote{The superhorizon modes, with \mbox{$k/{\cal H} \ll 1$}, are independent of the thermodynamic behavior. The density contrasts grow quadratically with $k/{\cal H}$ \cite{Schmid:1998mx}. }. This condition is equivalent to $q \,x\, \xi \gg 1$ or simply to $\xi \gg 1$. In Fig. \ref{fig:xi10ab} we show our results obtained for $\xi=10$ with the initial conditions: $\delta=1.0$, $\psi=\psi_{\rm ew}=0$, and $\delta_{\rm ew}=-0.5$ (upper part) and $\delta_{\rm ew}=0$ (lower part). The density contrast of the strongly interacting matter $\delta$ (solid line) and of the electro-weak sector $\delta_{\rm ew}$  (dashed line) oscillate with frequencies  determined by the value of the sound velocity in medium. This is indicated by the thin solid line representing the function
\begin{equation}
\cos\left( \xi \int\limits_0^x c_s^2(x^\prime) dx^\prime \right).
\label{g}
\end{equation}
The frequency of $\delta$ is reasonably well reproduced by Eq. (\ref{g}), however, in the considered time interval the amplitude of $\delta$ decreases by about 30\%. 

In Ref. \cite{Schmid:1998mx} a schematic crossover transition between the ideal quark-gluon plasma and a gas of massless hadrons was analyzed (in addition to the much more elaborated case of the first order phase transition). A similar damping was observed in this case, that was identified  with the reduction of the number of degrees of freedom during the crossover transition. During the first order phase transition this type of damping is also present but it merely reduces the huge amplification of the modes triggered by vanishing of the sound velocity.  

For the realistic QCD EOS the sound velocity does not drop to zero at $T_c$ and we deal only with the moderate damping of the inhomogeneities of strongly interacting matter. On the other hand, the inhomogeneities in the electro-weak sector, where the number of degrees of freedom is fixed, are not damped. They are well described by the function of the form (\ref{g}) with a suitably fixed phase and $c_s^2 = 1/3$ (not shown).

The lower part of Fig. \ref{fig:xi10ab} differs from the upper part by the initial value of $\delta_{\rm ew}$ that is set equal to zero in this case. Interestingly, the initial non-zero value of $\delta$ triggers oscillation in the electro-weak sector. Similarly to the upper part, the amplitude of the oscillations of $\delta$ is damped by about 30\% during the considered time interval. 

In Fig. \ref{fig:xi5ab} we show our results obtained for $\xi=5$. Similarly to the previous case with $\xi=10$ we observe again a slow damping of the oscillations of the QCD matter. 

\section{Conclusions}

In this paper we have considered the effect of the realistic QCD EOS on the evolution of the early Universe. We have compared the solutions of the Friedmann equation with those obtained with the bag EOS. The use of the realistic QCD EOS has small effects on the time evolution of the energy density, however, there are noticeable differences in the time evolution of pressure and temperature, which are more affected by the first order phase transition. Such difference may have an impact on various physical quantities if the cosmological measurements are done with high precision. The differences between the realistic QCD EOS and the bag EOS are the smallest if the critical temperature of the bag model is fitted to reproduce the critical temperature of the realistic crossover transition. 

In the second part of the paper we have considered the time development of the energy density fluctuations. We have shown that small inhomogeneities of the strongly interacting matter are moderately (by about 30\%) damped during the realistic crossover transition. Our results indicate that there are small chances for observation exotic phenomena connected with the first order phase transitions.

This research was supported in part by the Polish Ministry of Science and Higher Education, grant  No. N N202 263438.


\end{document}